\documentclass[authorversion,nonacm,screen,acmsmall]{acmart}

\AtBeginDocument{%
  \providecommand\BibTeX{{%
    \normalfont B\kern-0.5em{\scshape i\kern-0.25em b}\kern-0.8em\TeX}}}

\usepackage{subfigure}
\usepackage{todonotes}

\begin{document}

\title{A Survey of Privacy Attacks in Machine Learning}

\author{Maria Rigaki}
\orcid{0000-0002-0688-7752}
\email{maria.rigaki@fel.cvut.cz}
\affiliation{
  \institution{Czech Technical University in Prague}
  \streetaddress{Karlovo n\'{a}m\v{e}st\'{i} 13}
  \city{Prague}
  \state{Czech Republic}
  \postcode{120 00}
}

\author{Sebastian Garcia}
\email{sebastian.garcia@agents.fel.cvut.cz}
\affiliation{
  \institution{Czech Technical University in Prague}
  \streetaddress{Karlovo n\'{a}m\v{e}st\'{i} 13}
  \city{Prague}
  \state{Czech Republic}
  \postcode{120 00}
}

\renewcommand{\shortauthors}{Rigaki and Garcia}

\begin{abstract}
As machine learning becomes more widely used, the need to study its implications in security and privacy becomes more urgent. Although the body of work in privacy has been steadily growing over the past few years, research on the privacy aspects of machine learning has received less focus than the security aspects. Our contribution in this research is an analysis of more than 40 papers related to privacy attacks against machine learning that have been published during the past seven years. We propose an attack taxonomy, together with a threat model that allows the categorization of different attacks based on the adversarial knowledge, and the assets under attack. An initial exploration of the causes of privacy leaks is presented, as well as a detailed analysis of the different attacks. Finally, we present an overview of the most commonly proposed defenses and a discussion of the open problems and future directions identified during our analysis.
\end{abstract}



\maketitle

\section{Introduction}

Fueled by large amounts of available data and hardware advances, machine learning has experienced tremendous growth in academic research and real world applications. At the same time, the impact on the security, privacy, and fairness of machine learning is receiving increasing attention. In terms of privacy, our personal data are being harvested by almost every online service and are used to train models that power machine learning applications. However, it is not well known if and how these models reveal information about the data used for their training. If a model is trained using sensitive data such as location, health records, or identity information, then an attack that allows an adversary to extract this information from the model is highly undesirable. At the same time, if private data has been used without its owners' consent, the same type of attack could be used to determine the unauthorized use of data and thus work in favor of the user's privacy. 

Apart from the increasing interest on the attacks themselves, there is a growing interest in uncovering what causes privacy leaks and under which conditions a model is susceptible to different types of privacy-related attacks. There are multiple reasons why models leak information. Some of them are structural and have to do with the way models are constructed, while others are due to factors such as poor generalization or memorization of sensitive data samples. Training for adversarial robustness can also be a factor that affects the degree of information leakage.  

The focus of this survey is the privacy and confidentiality attacks on machine learning algorithms. That is, attacks that try to extract information about the training data or to extract the model itself. Some existing surveys~\cite{papernot2017practical, biggio2018wild} provide partial coverage of privacy attacks and there are a few other peer-reviewed works on the topic~\cite{jere2020ataxonomy, alrubaie2019privacypre}. However, these papers are either too high level or too specialized in a narrow subset of attacks.

The security of machine learning and the impact of adversarial attacks on the performance of the models have been widely studied in the community, with several surveys highlighting the major advances in the area~\cite{papernot2018sok,wang2019security, biggio2018wild, maiorca2019towards, liu2021privacysec}. Based on the taxonomy proposed in~\cite{biggio2018wild}, there are three types of attacks on machine learning systems: i) attacks against integrity, e.g., evasion and poisoning backdoor attacks that cause misclassification of specific samples, ii) attacks against a system's availability, such as poisoning attacks that try to maximize the misclassification error and iii) attacks against privacy and confidentiality, i.e., attacks that try to infer information about user data and models. While all attacks on machine learning are adversarial in nature, the term "adversarial attacks" is commonly used to refer to security-related attacks and more specifically to adversarial samples. In this survey, we only focus on privacy and confidentiality attacks.

An attack that extracts information about the model's structure and parameters is, strictly speaking, an attack against model confidentiality. The decision to include model extraction attacks was made because in the existing literature, attacks on model confidentiality are usually grouped together with privacy attacks~\cite{papernot2018sok, biggio2018wild}. Another important reason is that stealing model functionality may be considered a privacy breach as well. Veale et al.~\cite{veale2018algorithms} made the argument that privacy attacks such as membership inference (Section~\ref{subsec:mi_attacks}) increase the risk of machine learning models being classified as personal data under European Union's General Data Protection Regulation (GDPR) law because they can render a person identifiable. Although models are currently not covered by the GDPR, it may happen that they will be considered as personal data, and then attacks against them may fall under the same scope as attacks against personal data. This may be further complicated by the fact that model extraction attacks can be used as a stepping stone for other attacks. 

This paper is, as far as we know, the first \emph{comprehensive} survey of privacy-related attacks against machine learning. It reviews and systematically analyzes over 40 research papers. The papers have been published in top tier conferences and journals in the areas of security, privacy, and machine learning during 2014-2020. An initial set of papers was selected in Google Scholar using keyword searches related to "privacy", "machine learning", and the names of the attacks themselves ("membership inference", "model inversion", "property inference", model stealing", "model extraction", etc.). After the initial set of papers was selected, more papers were added by backward search based on their references as well as by forward search based on the papers that cited them.

The main contributions of this paper are:
\begin{itemize}
    \item The first comprehensive study of attacks on privacy and confidentiality of machine learning systems. 
    \item A unifying taxonomy of attacks against machine learning privacy.
    \item A discussion on the probable causes of privacy leaks in machine learning systems.
    \item An in-depth presentation of the implementation of the attacks.
    \item An overview of the different defensive measures tested to protect against the different attacks.
\end{itemize}

\subsection{Organization of the Paper}
The rest of the paper is organized as follows: Section 2 introduces some basic concepts related to machine learning that are relevant to the implementation of the attacks which are presented in Section~\ref{sec:attack_design}. The threat model is presented in Section~\ref{sec:threat_model} and the taxonomy of the attacks and their definition are the focus of Section~\ref{sec:taxonomy}. In Section~\ref{sec:why_leak} we present the causes of machine learning leaks that are known or have been investigated so far. An overview of the proposed defences per attack type is the focus of Section~\ref{sec:defenses}. Finally, Section~\ref{sec:discussion} contains a discussion on the current and future research directions and Section~\ref{sec:conclusion} offers concluding remarks.

\section{Machine Learning}
Machine learning (ML) is a field that studies the problem of learning from data without being explicitly programmed. The purpose of this section is to provide a non-exhaustive overview of machine learning as it pertains to this survey and to facilitate the discussion in the subsequent chapters. We briefly introduce a high level view of different machine learning paradigms and categorizations as well as machine learning architectures. Finally, we present a brief discussion on model training and inference. For the interested reader, there are several textbooks such as \cite{goodfellow2016deep, bishop2006pattern, murphy2012machine, shalev2014understanding} that provide a thorough coverage of the topic.  


\subsection{Types of Learning}
At a very high level, ML is traditionally split into three major areas: \textit{supervised}, \textit{unsupervised} and \textit{reinforcement} learning. Each of these areas has its own subdivisions. Over the years, new categories have emerged to capture types of learning that are not easily fit under these three areas such as \textit{semi-supervised} and \textit{self-supervised} learning, or other ways to categorize models such as \textit{generative} and \textit{discriminative} ones.


\subsubsection{Supervised Learning}
In a supervised learning setting, a model $f$ with parameters $\mathbf{\theta}$ is a mapping function between inputs $\mathbf{x}$ and outputs $\mathbf{y}=f(\mathbf{x};\mathbf{\theta})$, where $\mathbf{x}$ is a vector of attributes or features with dimensionality $n$. The output or label $\mathbf{y}$ can assume different dimensions depending on the learning task. A training set $\mathcal{D}$ used for training the model is a set of data points $\mathcal{D} = \{(\mathbf{x}_{i}, \mathbf{y}_{i})\}_{i=1}^{m}$, where $m$ is the number of input-output pairs. The most common supervised learning tasks are \textit{classification} and \textit{regression}. Examples of supervised learning algorithms include linear regression, logistic regression, decision trees, support vector machines, and many more. The vast majority of the attack papers thus far are focused in supervised learning using deep neural networks.

\subsubsection{Unsupervised Learning}
In unsupervised learning, there are no labels $\mathbf{y}$. The training set $\mathcal{D}$ consists only of the inputs $\mathbf{x}_{i}$. Unsupervised algorithms aim to find structure or patterns in the data without having access to labels. Usual tasks in unsupervised learning are \textit{clustering} \textit{feature learning}, \textit{anomaly detection} and \textit{dimensionality reduction}. In the context of this survey, attacks on unsupervised learning appear mostly as attacks on language models. 

\subsubsection{Reinforcement Learning}
Reinforcement learning concerns itself with agents that make observations of the environment and use these to take actions with the goal of maximizing a reward signal. In the most general formulation, the set of actions is not predefined and the rewards are not necessarily immediate but can occur after a sequence of actions~\cite{sutton2018reinforcement}. To our knowledge, no privacy-related attacks against reinforcement learning have been reported, but it has been used to launch other privacy-related attacks~\cite{orekondy2019knockoff}.

\subsubsection{Semi-supervised Learning} 
In many real-world settings, the amount of labeled data can be significantly smaller than that of unlabeled ones, and it might be too costly to obtain high-quality labels. Semi-supervised learning algorithms aim to use unlabeled data to learn higher level representations and then use the labeled examples to guide the downstream learning task. An example of semi-supervised learning would be to use an unsupervised learning technique such as clustering on unlabeled data and then use a classifier to separate representative training data from each cluster. Other notable examples are generative models such as Generative Adversarial Networks (GANs)~\cite{goodfellow2014gans}.


\subsubsection{Generative and Discriminative Learning}
Another categorization of learning algorithms is that of \textit{discriminative} vs \textit{generative} algorithms. Discriminative classifiers try to model the conditional probability $p(y | \mathbf{x})$, i.e., they try to learn the decision boundaries that separate the different classes directly based on the input data $\mathbf{x}$. Examples of such algorithms are logistic regression and neural networks. Generative classifiers try to capture the joint distribution $p(\mathbf{x}, y)$. An example of such a classifier is Naive Bayes. Generative models that do not require labels, but they try to model $p(\mathbf{x})$, explicitly or implicitly. Notable examples are language models that predict the next word(s) given some input text or GANs and Variational Autoencoders (VAEs)~\cite{diederik2014auto} that are able to generate data samples that match the properties of the training data.

\subsection{Learning Architectures}
From a system architecture point of view, we view the learning process as either a centralized or a distributed one. The main criterion behind this categorization is whether the data and the model are collocated or not. 

\subsubsection{Centralized Learning}
In a centralized learning setting, the data and the model are collocated. There can be one or multiple data producers or owners, but all data are gathered in one central place and used for the training of the model. The location of the data can be in a single or even multiple machines in the same data center. While using parallelism in the form of multiple GPUs and CPUs could be considered a distributed learning mode, it is not for us since we use the model and data collocation as the main criterion for the distinction between centralized and distributed learning. The centralized learning architecture includes the Machine Learning as a Service (MLaaS) setup, where the data owner uploads their data to a cloud-based service that is tasked with creating the best possible model. 

\subsubsection{Distributed Learning}
The requirements that drive the need for distributed learning architectures are the handling and processing of large amounts of data, the need for computing and memory capacity, and even privacy concerns. From the existing variants of distributed learning, we present those that are relevant from a privacy perspective, namely \textit{collaborative} or \textit{federated learning} (FL), \textit{fully decentralized} or \textit{peer-to-peer} (P2P) learning and \textit{split learning}.

Collaborative or federated learning is a form of decentralized training where the goal is to learn one global model from data stored in multiple remote devices or locations~\cite{li2019federated}. The main idea is that the data do not leave the remote devices. Data are processed locally and then used to update the local models. Intermediate model updates are sent to the central server that aggregates them and creates a global model. The central server then sends the global model back to all participant devices.

In fully decentralized learning or Peer-to-Peer (P2P) learning, there is no central orchestration server. Instead, the devices communicate in a P2P fashion and exchange their updates directly with other devices. This setup may be interesting from a privacy perspective, since it alleviates the need to trust a central server. However, attacks on P2P systems are relevant in such settings and need to be taken into account. Up to now, there were no privacy-based attacks reported on such systems; although they may become relevant in the future. Moreover, depending on the type of information shared between the peers, several of the attacks on collaborative learning may be applicable.

In split learning, the trained model is split into two or more parts. The edge devices keep the initial layers of the deep learning model and the centralized server keeps the final layers~\cite{hauswald2014hybrid, kang2017neuro}. The reason for the split is mainly to lower communication costs by sending intermediate model outputs instead of the input data. This setup is also relevant in situations where remote or edge devices have limited resources and are connected to a central cloud server. This scenario is common for Internet of Things (IoT) devices.

\subsection{Training and Inference}
Training of supervised ML models usually follows the Empirical Risk Minimization (ERM) approach~\cite{vapnik1992principles}, where the objective is to find the parameters $\mathbf{\theta}^*$ that minimize the \textit{risk} or \textit{objective function}, which is calculated as an average over the training dataset:
\begin{equation}
    \mathcal{J}(\mathcal{D};\mathbf{\theta}) = \frac{1}{m}\sum_{i=1}^{m}l(f(x_{i}; \mathbf{\theta}), y_{i})
\end{equation}

where $l(\cdot)$ is a loss function, e.g. cross entropy loss, and $m$ is the number of data points in the dataset $\mathcal{D}$. 

The idea behind ERM is that the training dataset is a subset drawn from the unknown true data distribution for the learning task. Since we have no knowledge of the true data distribution, we cannot minimize the true objective function, but instead we can minimize the estimated objective over the data samples that we have. In some cases, a regularization term is added to the objective function to reduce overfitting and stabilize the training process.

\subsubsection{Training in Centralized Settings}
The training process usually involves an iterative optimization algorithm such as gradient descent~\cite{cauchy1847methode}, which aims to minimize the objective function by following the path induced by its gradients. When the dataset is large, as is often the case with deep neural networks, taking one gradient step becomes too costly. In that case, variants of gradient descent which involve steps taken over smaller batches of data are preferred. One such optimization method is called Stochastic Gradient Descent (SGD)~\cite{robbins1951stochastic} defined by:
\begin{equation}
        \label{eq:theta_update}
        \mathbf{\theta}_{t+1} = \mathbf{\theta}_{t} - \eta \textbf{g}  
\end{equation}

\begin{equation}
    \label{eq:sgd_grad}
     \textbf{g} = \frac{1}{m'} \nabla_{\theta} \sum_{i=1}^{m'}  l(f(\mathbf{x}_{i}; \mathbf{\theta}), \mathbf{y}_{i})
\end{equation}

where $\eta$ is the learning rate and $\mathbf{g}$ is the gradient of the loss function with respect to parameters $\mathbf{\theta}$. In the original formulation of SGD the gradient $\mathbf{g}$ is calculated over a single data point from $\mathcal{D}$, chosen randomly, hence the name stochastic. In practice, it is  common to use mini-batches of size $m'$ where $m' < m$, instead of a single data point to calculate the loss gradient at each step (Equation~\ref{eq:sgd_grad}). Mini-batches lower the variance of the stochastic gradient estimate, but the size $m'$ is a tunable parameter that can affect the performance of the algorithm. While SGD is still quite popular, several improvements have been proposed to try to speed up convergence by adding momentum~\cite{polyak1964some}, by using adaptive learning rates as, for example, in the  RMSprop algorithm~\cite{hinton2012neural}, or by combining both improvements as in the Adam algorithm~\cite{kingma2014adam}.

\subsubsection{Training in Distributed Settings}

The most popular learning algorithm for federated learning is federated averaging~\cite{pmlr-v54-mcmahan17a}, where each remote device calculates one step of gradient descent from the locally stored data and then shares the updated model weights with the parameter server. The parameter server averages the weights of all remote participants and updates the global model which is subsequently shared again with the remote devices. It can be defined by:

\begin{equation}
    \label{eq:fed_avg}
    \theta_{t+1} = \frac{1}{K} \sum_{k=1}^{K} \theta_{t}^{(k)} 
\end{equation}

where K is the number of remote participants and the parameters $\theta_{t}^{(k)}$ of participant $k$ have been calculated locally based on Equations~\ref{eq:theta_update} and \ref{eq:sgd_grad}. 

Another approach that comes from the area of distributed computing is downpour (or synchronized) SGD~\cite{dean2012large}, which proposes to share the loss gradients of the distributed devices with the parameter server that aggregates them and then performs one step of gradient descent. It can be defined by:

\begin{equation}
    \label{eq:SSGD}
    \theta_{t+1} = \theta_{t} - \eta \sum_{k=1}^{K} \frac{m^{(k)}}{M}\mathbf{g}_t^{(k)}
\end{equation}

where $\mathbf{g}_t^{(k)}$ is the gradient computed by participant $k$ based on Equation~\ref{eq:sgd_grad} using their local data, $m^{(k)}$ is the number of data points in the remote participant and $M$ is the total number of data points in the training data. After the calculation of Equation~\ref{eq:SSGD}, the parameter server sends the updated model parameters $\theta_{t+1}$ to the remote participants.

\subsubsection{Inference}
Once the models are trained, they can be used to make inferences or predictions over previously unseen data. At this stage, the assumption is that the model parameters are fixed, although the models are usually monitored, evaluated, and retrained if necessary. The majority of the attacks in this survey are attacks during the inference phase of the model lifecycle except for the attacks on collaborative learning which are usually performed during training.

\section{Threat Model}
\label{sec:threat_model}
To understand and defend against attacks in machine learning from a privacy perspective, it is useful to have a general model of the environment, the different actors, and the assets to protect. 

From a threat model perspective, the assets that are sensitive and are potentially under attack are the training dataset $\mathcal{D}$, the model itself, its parameters $\mathbf{\theta}$, its hyper-parameters, and its architecture. The actors identified in this threat model are: 
\begin{enumerate}
    \item The \textbf{data owners}, whose data may be sensitive.
    \item The \textbf{model owners}, which may or may not own the data and may or may not want to share information about their models.
    \item The \textbf{model consumers}, that use the services that the model owner exposes, usually via some sort of programming or user interface.
    \item The \textbf{adversaries}, that may also have access to the model's interfaces as a normal consumer does. If the model owner allows, they may have access to the model itself. 
\end{enumerate}

\begin{figure}[t]
\centering
\includegraphics[width=10cm]{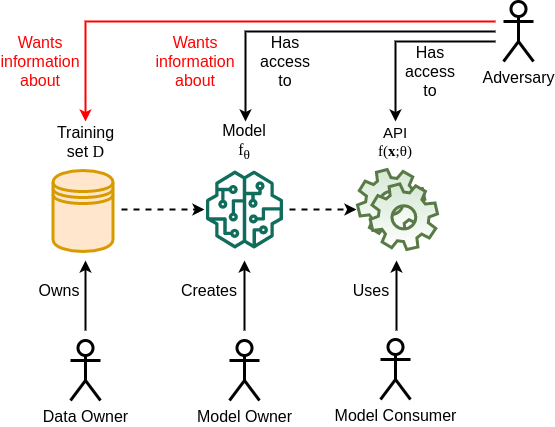}
\caption{Threat Model of privacy and confidentiality attacks against machine learning systems. The human figure represents actors and the symbols represent the assets. Dashed lines represent data and information flow, while full lines represent possible actions. In red are the actions of the adversaries, available under the threat model.}
\label{fig:threat_model}
\end{figure}

Figure~\ref{fig:threat_model} depicts the assets and the identified actors under the threat model, as well as the information flow and possible actions. This threat model is a logical model and it does not preclude the possibility that some of these assets may be collocated or spread in multiple locations.

\begin{figure*}[t]
\centering
\includegraphics[width=9cm]{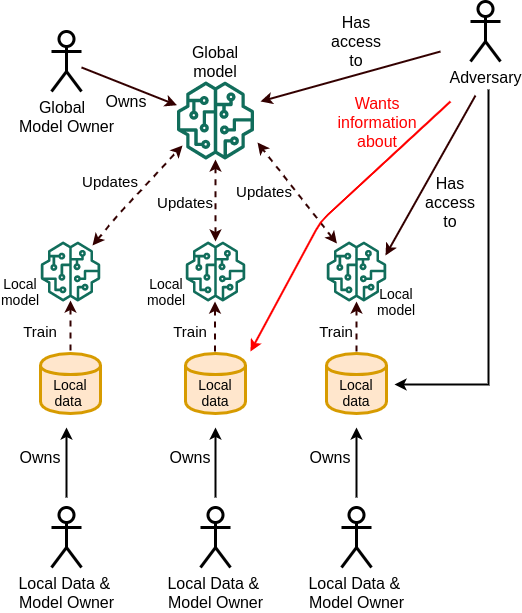}
\caption{Threat model in a collaborative learning setting. Dashed lines represent data and information flows, while full lines represent possible actions. In red are the actions of the adversaries, available under the threat model. In this setting the adversary can be placed either at the parameter server or locally. Model consumers are not depicted for reasons of simplicity. In a federated learning setting, local model owners are also model consumers.}
\label{fig:federated_threat}
\end{figure*}

Distributed modes of learning, such as federated or collaborative learning, introduce different spatial models of adversaries. In a federated learning setting, the adversary can be collocated with the global model, but it can also be a local attacker. Figure~\ref{fig:federated_threat} shows the threat model in a collaborative learning setting. The presence of multiple actors allows also the possibility of \textit{colluding} adversaries that join forces.

The different attack surfaces against machine learning models can be modelled in terms of \textbf{adversarial knowledge}. The range of knowledge varies from limited, e.g., having access to a machine learning API, to having knowledge of the full model parameters and training settings. In between these two extremes, there is a range of possibilities such as partial knowledge of the model architecture, its hyper-parameters, or training setup. The knowledge of the adversary can also be considered from a dataset point of view. In the majority of the papers reviewed, the authors assume that the adversaries have no knowledge of the training data samples, but they may have some knowledge of the underlying data distribution. 

From a taxonomy point of view, attacks where the adversary has no knowledge of the model parameters, architecture, or training data are called \textbf{black-box} attacks. An example of a black-box system is Machine Learning as a Service (MLaaS) where the users usually provide some input and receive either a prediction vector or a class label from a pre-trained model hosted in the cloud. Most black-box papers assume the existence of a prediction vector. In a similar fashion, \textbf{white-box} attacks are those where the adversary has either complete access to the target model parameters or their loss gradients during training. This is the case, for example, in most distributed modes of training. In between the two extremes, there are also attacks that make stronger assumptions than the black-box ones, but do not assume full access to the model parameters. We refer to these attacks as \textbf{partial white-box} attacks. It is important to add here that the majority of the works assume full knowledge of the expected input, although some form of preprocessing might be required.

The time of the attack is another parameter to consider from a taxonomy point of view. The majority of the research in the area is dealing with attacks during \textbf{inference}, however most collaborative learning attacks assume access to the model parameters or gradients during \textbf{training}. Attacks during the training phase of the model open up the possibility for different types of adversarial behavior. A \textbf{passive} or \textit{honest-but-curious} attacker does not interfere with the training process and they are only trying to infer knowledge during or after the training. If the adversary interferes with the training in any way, they are considered an \textbf{active} attacker. 

Finally, since the interest of this survey is in privacy attacks based on unintentional information leakage regarding the data or the machine learning model, there is no coverage of \textit{security-based} attacks, such as model poisoning or evasion attacks, or attacks against the infrastructure that hosts the data, models or provided services.

\section{Attack Types}
\label{sec:taxonomy}
In privacy-related attacks, the goal of an adversary is to gain knowledge that was not intended to be shared. Such knowledge can be about the training data $\mathcal{D}$ or information about the model, or even extracting information about properties of the data such as unintentionally encoded biases. In our taxonomy, the privacy attacks studied are categorized into four types: \textbf{membership inference}, \textbf{reconstruction}, \textbf{property inference}, and \textbf{model extraction}.

\subsection{Membership Inference Attacks}
\label{subsec:mi_attacks}
Membership inference tries to determine whether an input sample $\mathbf{x}$ was used as part of the training set $\mathcal{D}$. This is the most popular category of attacks and was first introduced by Shokri et al.~\cite{shokri2017membership}. The attack only assumes knowledge of the model's output prediction vector (black-box) and was carried out against supervised machine learning models. White-box attacks in this category are also a threat, especially in a collaborative setting, where an adversary can mount both passive and active attacks. If there is access to the model parameters and gradients, then this allows for more effective white-box membership inference attacks in terms of accuracy~\cite{nasr2019comprehensive}. 

Apart from supervised models, generative models such as GANs and VAEs are also susceptible to membership inference attacks~\cite{hayes2019logan, hilprecht2019monte, chen2019gan}. The goal of the attack, in this case, is to retrieve information about the training data using varying degrees of knowledge of the data generating components.

Finally, these types of attacks can be viewed from a different perspective, that of the data owner. In such a scenario, the owner of the data may have the ability to audit black-box models to see if the data have been used without authorization~\cite{song2019auditing, hishamoto2020embership}.   

\subsection{Reconstruction Attacks}

Reconstruction attacks try to recreate one or more training samples and/or their respective training labels. The reconstruction can be partial or full. Previous works have also used the terms \textbf{attribute inference} or \textbf{model inversion} to describe attacks that, given output labels and partial knowledge of some features, try to recover sensitive features or the full data sample. For the purpose of this survey, all these attacks are considered as part of the larger set of reconstruction attacks. The term \textbf{attribute inference} has been used in other parts of the privacy related literature to describe attacks that infer sensitive "attributes" of a targeted user by leveraging publicly accessible data~\cite{gong2016you, jia2018attriguard}. These attacks are not part of this review as they are mounted against the individual's data directly and not against ML models.

A major distinction between the works of this category is between those that create an actual reconstruction of the data~\cite{zhu2019dlg, he2019collaborative, wang2019beyondclass, yang2019neural, zhang2020secret} and the ones that create class representatives or probable values of sensitive features that do not necessarily belong to the training dataset~\cite{fredrikson2014pharma, hitaj2017deep, yang2019neural, hidano2017model}. In classification models, the latter case is limited to scenarios where classes are made up of one type of object, e.g., faces of the same person. While this limits the applicability of the attack, it can still be an interesting scenario in some cases.

\subsection{Property Inference Attacks}
The ability to extract dataset properties which were not explicitly encoded as features or were not correlated to the learning task, is called \textbf{property inference}. An example of property inference is the extraction of information about the ratio of women and men in a patient dataset when this information was not an encoded attribute or a label of the dataset. Or having a neural network that performs gender classification and can be used to infer if people in the training dataset wear glasses or not. In some settings, this type of leak can have privacy implications. These types of properties can also be used to get more insight about the training data, which can lead to adversaries using this information to create similar models \cite{ateniese2015hacking} or even have security implications when the learned property can be used to detect vulnerabilities of a system~\cite{ganju2018property}. 

Property inference aims to extract information that was learned from the model unintentionally and that is not related to the training task. Even well generalized models may learn properties that are relevant to the whole input data distribution and sometimes this is unavoidable or even necessary for the learning process. What is more interesting from an adversarial perspective, are properties that may be inferred from a specific subset of training data, or eventually about a specific individual.

Property inference attacks so far target either dataset-wide properties~\cite{ateniese2015hacking, ganju2018property, song2020infoleak} or the emergence of properties within a batch of data~\cite{melis2019exploiting}. The latter attack was performed on the collaborative training of a model.

\subsection{Model Extraction Attacks}
\label{subsec:model_extraction}
\textbf{Model extraction} is a class of black-box attacks where the adversary tries to extract information and potentially fully reconstruct a model by creating a substitute model $\hat{f}$ that behaves very similarly to the model under attack $f$. There are two main focus for substitute models. First, to create models that match the accuracy of the target model $f$ in a test set that is drawn from the input data distribution and related to the learning task~\cite{milli2019model, orekondy2019knockoff, tramer2016stealing, krishna2020Thieves}. Second, to create a substitute model $\hat{f}$ that matches $f$ at a set of input points that are not necessarily related to the learning task~\cite{tramer2016stealing, juuti2019prada, Correia-Silva-IJCNN2018, jagielski2020high}. Jagielski et al.~\cite{jagielski2020high} referred to the former attack as \textbf{task accuracy} extraction and the latter as \textbf{fidelity} extraction. In task accuracy extraction, the adversary is interested in creating a substitute that learns the same task as the target model equally well or better. In the latter case, the adversary aims to create a substitute that replicates the decision boundary of $f$ as faithfully as possible. This type of attack can be later used as a stepping stone before launching other types of attacks such as adversarial attacks~\cite{papernot2017practical, juuti2019prada} or membership inference attacks~\cite{nasr2019comprehensive}. In both cases, it is assumed that the adversary wants to be as efficient as possible, i.e., to use as few queries as possible. Knowledge of the target model architecture is assumed in some works, but it is not strictly necessary if the adversary selects a substitute model that has the same or higher complexity than the model under attack~\cite{orekondy2019knockoff, juuti2019prada, krishna2020Thieves}. 

Apart from creating substitute models, there are also approaches that focus on recovering information from the target model, such as hyper-parameters in the objective function~\cite{wang2018stealing} or information about various neural network architectural properties such as activation types, optimisation algorithm, number of layers, etc~\cite{joon2018towards}.

\section{Causes of Privacy Leaks} 
\label{sec:why_leak}

The conditions under which machine learning models leak is a research topic that has started to emerge in the past few years. Some models leak information due to the way they are constructed. An example of such a case is Support Vector Machines (SVMs), where the support vectors are data points from the training dataset. Other models, such as linear classifiers are relatively easy to "reverse engineer" and to retrieve their parameters just by having enough input / output data pairs~\cite{tramer2016stealing}. Larger models such as deep neural networks usually have a large number of parameters and simple attacks are not feasible. However, under certain assumptions and conditions, it is possible to retrieve information about either the training data or the models themselves. 


\subsection{Causes of Membership Inference Attacks}
One of the conditions that has been shown to improve the accuracy of membership inference is the poor generalization of the model. The connection between overfitting and black-box membership inference was initially investigated by Shokri et al.~\cite{shokri2017membership}. This paper was the first to examine membership inference attacks on neural networks. The authors measured the effect of overfitting on the attack accuracy by training models in different MLaaS platforms using the same dataset. The authors showed experimentally that overfitting can lead to privacy leakage but also noted that it is not the only condition, since some models that had lower generalization error where more prone to membership leaks. The effect of overfitting was later corroborated formally by Yeom et al.~\cite{yeom2018privacy}. The authors defined membership advantage as a measure of how well an attacker can distinguish whether a data sample belongs to the training set or not, given access to the model. They proved that the membership advantage is proportional to the generalization error of the model and that overfitting is a sufficient condition for performing membership inference attacks but not a necessary one. Additionally, Long et al.~\cite{long2018understanding} showed that even in well-generalized models, it is possible to perform membership inference for a subset of the training data which they named \textit{vulnerable records}. 

Other factors, such as the model architecture, model type, and dataset structure, affect the attack accuracy. Similarly to ~\cite{shokri2017membership} but in the white-box setting, Nasr et al.~\cite{nasr2019comprehensive} showed that two models with the same generalization error showed different degrees of leakage. More specifically, the most complex model in terms of number of parameters exhibited higher attack accuracy, showing that model complexity is also an important factor. 

Truex et al.~\cite{truex2019demystifying} ran different types of experiments to measure the significance of the model type as well as the the number of classes present in the dataset. They found that certain model types such as Naive Bayes are less susceptible to membership inference attacks than decision trees or neural networks. They also showed that as the number of classes in the dataset increases, so does the potential of membership leaks. This finding agrees with the results in ~\cite{shokri2017membership}.

Securing machine learning models against adversarial attacks can also have an adverse effect on the model's privacy as shown by Song et al.~\cite{song2019privacy}. Current state of the art proposals for robust model training, such as projective gradient descent (PGD) adversarial training~\cite{madry2018towards}, increase the model's susceptibility to membership inference attacks. This is not unexpected since robust training methods (both empirical and provable defenses) tend to increase the generalization error. As previously discussed, the generalization error is related to the success of the attack. Furthermore, the authors of~\cite{song2019privacy} argue that robust training may lead to increased model sensitivity to the training data, which can also affect membership inference.

The generalization error is easily measurable in supervised learning under the assumption that the test data can capture the nuances of the real data distribution. In generative models and specifically in GANs this is not the case, hence the notion of overfitting is not directly applicable. All three papers that deal with membership inference attacks against GANs mention overfitting as an important factor behind successful attacks~\cite{hayes2019logan, hilprecht2019monte, chen2019gan}. In this case, overfitting means that the generator has memorized and replays part of the training data. This is further corroborated in the study in~\cite{chen2019gan}, where their attacks are shown to be less successful as the training data size increases.

\subsection{Causes of Reconstruction Attacks}
Regarding reconstruction attacks, Yeom et al.~\cite{yeom2018privacy} showed that a higher generalization error can lead to a higher probability to infer data attributes, but also that the influence of the target feature on the model is an important factor. However, the authors assumed that the adversary has knowledge of the prior distribution of the target features and labels. Using weaker assumptions about the adversary's knowledge, Zhang et al.~\cite{zhang2020secret} showed theoretically and experimentally that a model that has high predictive power is more susceptible to reconstruction attacks. Finally, similarly to vulnerable records in membership inference, memorization and retrieval of data which are \textit{out-of-distribution} was shown to be the case even for models that do not overfit~\cite{carlini2019secret}.


\subsection{Causes of Property Inference Attacks}

Property inference is possible even with well-generalized models~\cite{melis2019exploiting, ganju2018property} so overfitting does not seem to be a cause of property inference attacks. Unfortunately, regarding property inference attacks, we have less information about what makes them possible and under which circumstances they appear to be effective. This is an interesting avenue for future research, both from a theoretical and an empirical point of view.

\subsection{Causes of Model Extraction}

While overfitting increases the success of black-box membership inference attacks, the exact opposite holds for model extraction attacks. It is possible to steal model parameters when the models under attack have 98\% or higher accuracy in the test set~\cite{joon2018towards}. Also models with a higher generalization error are harder to steal, probably due to the fact that they may have memorized samples that are not part of the attacker's dataset~\cite{liu2021ml}. Another factor that may affect model extraction success is the dataset used for training. Higher number of classes may lead to worse attack performance~\cite{liu2021ml}.


\section{Implementation of the Attacks}
\label{sec:attack_design}

More than 40 papers were analyzed in relation to privacy attacks against machine learning. This section describes in some detail the most commonly used techniques as well as the essential differences between them. The papers are discussed in two sections: attacks on centralized learning and attacks on distributed learning.

\subsection{Attacks Against Centralized Learning}
In the centralized learning setting, the main assumption is that models and data are collocated during the training phase. The next subsection introduces a common design approach that is used by multiple papers, namely, the use of \textit{shadow models} or \textit{shadow training}. The rest of the subsections are dedicated to the different attack types and introduce the assumptions, common elements as well as differences of the reviewed papers.

\subsubsection{Shadow training}
\begin{figure*}[t]
\centering
\includegraphics[width=12cm]{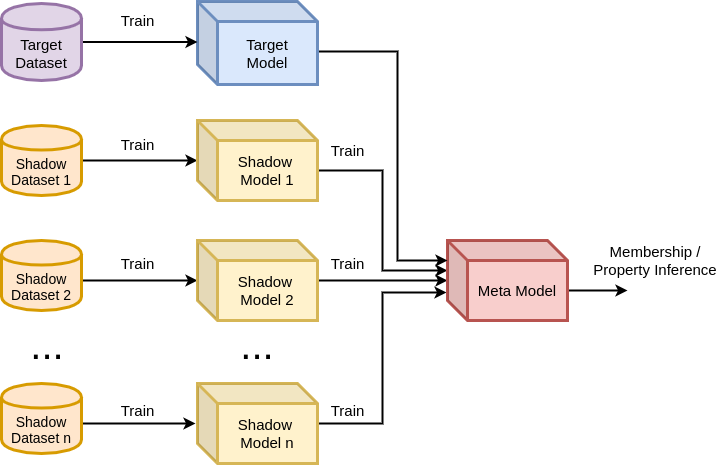}
\caption{Shadow training architecture. At first, a number of shadow models are trained with their respective shadow datasets in order to emulate the behavior of the target model. At the second stage, a meta-model is being trained from the outputs of the shadow models and the known labels of the shadow datasets. The meta-model is used to infer membership or properties of data or the model given the output of the target model.}
\label{fig:shadow}
\end{figure*}

A common design pattern for a lot of supervised learning attacks is the use of \textbf{shadow models} and \textbf{meta-models} or \textbf{attack-models}~\cite{ateniese2015hacking, shokri2017membership, ganju2018property, joon2018towards, rahman2018membership, jayaraman2019evaluating, Salem0HBF019, truex2019demystifying, sablayrolles2019plmr, hishamoto2020embership}. The general shadow training architecture is depicted in Figure~\ref{fig:shadow}. The main intuition behind this design is that models behave differently when they see data that do not belong to the training dataset. This difference is captured in the model outputs as well as in their internal representations. In most designs there is a target model and a target dataset. The adversary is trying to infer either membership or properties of the training data. They train a number of shadow models using shadow datasets $\mathcal{D}_{shadow} = \{\textbf{x}_{shadow,i}, \textbf{y}_{shadow,i}\}_{i=1}^{n}$ that usually are assumed to come from the same distribution as the target dataset. After the shadow models' training, the adversary constructs an attack dataset  $\mathcal{D}_{attack}=\{f_i(\textbf{x}_{shadow, i}), \textbf{y}_{shadow, i}\}_{i=1}^{n}$, where $f_i$ is the respective shadow model. The attack dataset is used to train the meta-model, which essentially performs inference based on the outputs of the shadow models. Once the meta-model is trained, it is used for testing using the outputs of the target model.

\subsubsection{Membership inference attacks} 

In \textit{membership inference} black-box attacks, the most common attack pattern is the use of shadow models. The output of the shadow models is usually a prediction vector~\cite{shokri2017membership, Salem0HBF019, rahman2018membership, truex2019demystifying, jayaraman2019evaluating}.  The labels used for the attack dataset come from the test and training splits of the shadow data, where the data points that belong to the test set are labeled as non-members of the training set. The meta-model is trained to recognize patterns in the prediction vector output of the target model. These patterns allow the meta-model to infer whether a data point belongs to the training dataset or not. The number of shadow models affects the attack accuracy, but it also incurs cost to the attackers. Salem et al.~~\cite{Salem0HBF019} showed that membership inference attacks are possible with as little as one shadow model. 

Shadow training can be further reduced to a threshold-based attack, where instead of training a meta-model, one can calculate a suitable threshold function that indicates whether a sample is a member of the training set. The threshold can be learned from multiple shadow models~\cite{sablayrolles2019plmr} or even without using any shadow models~\cite{yeom2018privacy}. Sablayrolles et al.~\cite{sablayrolles2019plmr} showed that a Bayes optimal membership inference attack depends only on the loss and their attack outperforms previous attacks such as~\cite{shokri2017membership, yeom2018privacy}. In terms of attack accuracy, they reported up to 90.8\% on large neural network models such as VGG16~\cite{liu2015very} that were performing classification on the Imagenet~\cite{imagenet_cvpr09} dataset.

In addition to relaxations on the number of shadow models, attacks have been shown to be data driven, i.e., an attack can be successful even if the target model is different than the shadow and meta-models~\cite{truex2019demystifying}. The authors tested several types of models such as k-NN, logistic regression, decision trees and naive Bayes classifiers in different combinations on the role of the target model, shadow and meta model. The results showed that i) using different types of models did not affect the attack accuracy and ii) in most cases, models such as decision trees outperformed neural networks in terms of attack accuracy and precision.

Shadow model training requires a shadow dataset. One of the main assumptions of membership inference attacks on supervised learning models is that the adversary has no or limited knowledge of the training samples used. However, the adversary knows something about the underlying data distribution of the training data. If the adversary does not have access to a suitable dataset, they can try to generate one~\cite{shokri2017membership, truex2019demystifying}. Access to statistics about the probability distribution of several features allows an attacker to create the shadow dataset using sampling techniques. If a statistics-based generation is not possible, a query-based approach using the target models' prediction vectors is another possibility. Generating auxiliary data using GANs was also proposed by Hayes et al.~\cite{hayes2019logan}. If the adversary manages to find input data that generate predictions with high confidence, then no prior knowledge of the data distribution is required for a successful attack~\cite{shokri2017membership}. Salem et al.~\cite{Salem0HBF019} went so far as to show that it is not even necessary to train the shadow models using data from the same distribution as the target, making the attack more realistic since it does not assume any knowledge of the training data.    

The previous discussion is mostly relevant to supervised classification or regression tasks. The efficacy of membership inference attacks against sequence-to-sequence models training for machine translation, was studied by~\cite{hishamoto2020embership}. The authors used shadow models that try to mimic the target model's behavior and then used a meta-model to infer membership. They found that sequence generation models are much harder to attack compared to other types of models such as image classification. However, membership of \textit{out-of-domain} and \text{out-of-vocabulary} data was easier to infer. 

Membership inference attacks are also applicable to deep generative models such as GANs and VAEs~\cite{hayes2019logan, hilprecht2019monte, chen2019gan}. Since these models have more than one component (generator/discriminator, encoder/decoder), adversarial knowledge needs to take that into account. For these types of models, the taxonomy proposed by Chen et al.~\cite{chen2019gan} is partially followed. We consider black-box access to the generator as the ability to access generated samples and partial black-box access, the ability to provide inputs $z$ and generate samples. Having access to the generator model and its parameters is considered a white-box attack. The ability to query the discriminator is also  a white-box attack. 

The full white-box attacks with access to the GAN discriminator are based on the assumption that if the GAN has "overfitted", then the data points used for its training will receive higher confidence values as output by the discriminator~\cite{hayes2019logan}. In addition to the previous attack, Hayes et al.~\cite{hayes2019logan} proposed a set of attacks in the partial black-box setting. These attacks are applicable to both GANs and VAEs or any generative model. If the adversary has no auxiliary data, they can attempt to train an auxiliary GAN  whose discriminator distinguishes between the data generated by the target generator and the data generated by the auxiliary GAN. Once the auxiliary GAN is trained, its discriminator can be used for the white-box attack. The authors considered also scenarios where the adversary may have auxiliary information such as knowledge of training and test data. Using the auxiliary data, they can train another GAN whose discriminator would be able to distinguish between members of the original training set and non-members.  


A distance-based attack over the nearest neighbors of a data point was proposed by Chen et al.~\cite{chen2019gan} for the full black-box model. In this case, a data point $\mathbf{x}$ is a member of the training set if within its k-nearest neighbors there is at least one point that has a distance lower than a threshold $\epsilon$. The authors proposed more complex attacks as the level of knowledge of the adversary increases, based on the idea that the reconstruction error between the real data point $x$ and a sample generated by the generator given some input $z$ should be smaller if the data point is coming from the training set. 




\subsubsection{Reconstruction attacks}

The initial reconstruction attacks were based on the assumption that the adversary has access to the model $f$, the priors of the sensitive and nonsensitive features, and the output of the model for a specific input $x$. The attack was based on estimating the values of sensitive features, given the values of nonsensitive features and the output label~\cite{fredrikson2014pharma}. This method used a maximum a posteriori (MAP) estimate of the attribute that maximizes the probability of observing the known parameters.
Hidano et al.~\cite{hidano2017model} used a similar attack but they made no assumption about the knowledge of the nonsensitive attributes. In order for their attack to work, they assumed that the adversary can perform a \textit{model poisoning} attack during training.

Both previous attacks worked against linear regression models, but as the number of features and their range increases, the attack feasibility decreases. To overcome the limitations of the MAP attack, Fredrikson et al.~\cite{fredrikson2015model} proposed another inversion attack which recovers features using target labels and optional auxiliary information. The attack was formulated as an optimization problem where the objective function is based on the observed model output and uses gradient descent in the input space to recover the input data point. The method was tested on image reconstruction. The result was a class representative image which in some cases was quite blurry even after denoising. A formalization of the model inversion attacks in \cite{fredrikson2014pharma, fredrikson2015model} was later proposed by Wu et al.~\cite{wu2016methodology}. 

Since the optimization problem in~\cite{fredrikson2015model} is quite hard to solve, Zhang et al.~\cite{zhang2020secret} proposed to use a GAN to learn some auxiliary information of the training data and produce better results. The auxiliary information in this case is the presence of blurring or masks in the input images. The attack first uses the GAN to learn to generate realistic looking images from masked or blurry images using public data. The second step is a GAN inversion that calculates the latent vector $\hat{z}$ which generates the most likely image: 

\begin{equation}
  \hat{z}=\arg\min_z L_{prior}(z) + \lambda L_{id}(z)  
\end{equation}

where the prior loss $L_{prior}$ is ensuring the generation of realistic images and $L_{id}$ ensures that the images have a high likelihood in the target network. The attack is quite successful, especially on masked images.

The only black-box reconstruction attack until now was proposed by Yang et al.~\cite{yang2019neural}. This attack employs an additional classifier that performs an inversion from the output of the target model $f(x)$ to a candidate output $\hat{x}$. The setup is similar to that of an autoencoder, only in this case the target network that plays the role of the encoder is a black box and it is not trainable. The attack was tested on different types of target model outputs: the full prediction vector, a truncated vector, and the target label only. When the full prediction vector is available, the attack performs a good reconstruction, but with less available information, the produced data point looks more like a class representative.

\subsubsection{Property inference attacks}

In \textit{property inference} the shadow datasets are labeled based on the properties that the adversary wants to infer, so the adversary needs access to data that have the property and data that do not have it. The meta-model is then trained to infer differences in the output vectors of the data that have the property versus the ones that they do not have it. In white-box attacks, the meta-model input can be other feature representations such as the support vectors of an SVM~\cite{ateniese2015hacking} or transformations of neural network layer outputs~\cite{ganju2018property}. When attacking language model embeddings, the embedding vectors themselves can be used to train a classifier to distinguish between properties such as text authorship~\cite{song2020infoleak}.

\subsubsection{Model extraction attacks}
When the adversary has access to the inputs and prediction outputs of a model, it is possible to view these pairs of inputs and outputs as a system of equations, where the unknowns are the model parameters~\cite{tramer2016stealing} or hyper-parameters of the objective function~\cite{wang2018stealing}. In the case of a linear binary classifier, the system of equations is linear and only $d + 1$ queries are necessary to retrieve the model parameters, where $d$ is the dimension of the parameter vector $\theta$. In more complex cases, such as multi-class linear regression or multi-layer perceptrons, the systems of equations are no longer linear. Optimization techniques such as Broyden–Fletcher–Goldfarb–Shanno (BFGS)~\cite{nocedal2006numerical} or stochastic gradient descent are then used to approximate the model parameters~\cite{tramer2016stealing}. 

Lack of prediction vectors or a high number of model parameters renders equation solving attacks inefficient. A strategy is required to select the inputs that will provide the most useful information for model extraction. From this perspective, model extraction is quite similar to \textit{active learning}~\cite{chandrasekaran2020exploring}. Active learning makes use of an external 
oracle that provides labels to input queries. The oracle can be a human expert or a system. The labels are then used to train or update the model. In the case of model extraction, the target model plays the role of the oracle.

Following the active learning approach, several papers propose an adaptive training strategy. They start with some initial data points or \textit{seeds} which they use to query the target model and retrieve labels or prediction vectors which they use to train the substitute model $\hat{f}$. For a number of subsequent rounds, they extend their dataset with new synthetic data points based on some adaptive strategy that allows them to find points close to the decision boundary of the target model~\cite{tramer2016stealing, juuti2019prada, papernot2017practical, chandrasekaran2020exploring}. Chandrasekaran et al.~\cite{chandrasekaran2020exploring} provided a more query efficient method of extracting nonlinear models such as kernel SVMs, with slightly lower accuracy than the method proposed by Tramer et al.~\cite{tramer2016stealing}, while the opposite was true for Decision Tree models.

Several other strategies for selecting the most suitable data for querying the target model use: (i) data that are not synthetic but belong to different domains such as images from different datasets~\cite{Correia-Silva-IJCNN2018, orekondy2019knockoff, barbalau2020ripper}, (ii) semi-supervised learning techniques such as rotation loss~\cite{zhai2019s4l} or MixMatch~\cite{berthelot2019mixmatch} to augment the dataset~\cite{jagielski2020high} or (iii) randomly generated input data~\cite{tramer2016stealing, juuti2019prada, krishna2020Thieves}. In terms of efficiency, semi-supervised methods such as MixMatch require much fewer queries than fully supervised extraction methods to perform similarly or better in terms of task accuracy and fidelity, against models trained for classification using CIFAR-10 and SVHN datasets~\cite{jagielski2020high}. For larger models, trained for Imagenet classification, even querying a 10\% of the Imagenet data, gives a comparable performance to the target model~\cite{jagielski2020high}. Against a deployed MLaaS service that provides facial characteristics, Orekondy et al.~\cite{orekondy2019knockoff} managed to create a substitute model that performs at 80\% of the target in task accuracy, spending as little as \$30.

Some, mostly theoretical, work has demonstrated the ability to perform direct model extraction beyond linear models~\cite{milli2019model, jagielski2020high}. Full model extraction was shown to be theoretically possible against two-layer fully connected neural networks with rectified linear unit (ReLU) activations by Milli et al.~\cite{milli2019model}. However, their assumption was that the attacker has access to the loss gradients with respect to the inputs. Jagielski et al.~\cite{jagielski2020high} managed to do a full extraction of a similar network without the need of gradients. Both approaches take into account that ReLUs transforms the neural network into a piecewise linear function of the inputs. By probing the model with different inputs, it is possible to identify where the linearity breaks and use this knowledge to calculate the network parameters. In a hybrid approach that uses both a learning strategy and direct extraction, Jagielski et al.~\cite{jagielski2020high}, showed that they can extract a model trained on MNIST with almost 100\% fidelity by using an average of $2^{19.2}$ to $2^{22.2}$ queries against models that contain up to 400,000 parameters. However, this attack assumes access to the loss gradients similarly to~\cite{milli2019model}.

Finally, apart from learning substitute models directly, there is also the possibility of extracting model information such as architecture, optimization methods and hyper-parameters using shadow models~\cite{joon2018towards}. The majority of attacks were performed against neural networks trained on MNIST. Using the shadow models' prediction vectors as input, the meta-models managed to learn to distinguish whether a model has certain architectural properties. An additional attack by the same authors, proposed to generate adversarial samples which were created by models that have the property in question. The generated samples were created in a way that makes a classifier output a certain prediction if they have the attribute in question. The target model's prediction on this adversarial sample is then used to establish if the target model has a specific property. The combination of the two attacks proved to be the most effective approach. Some properties such as activation function, presence of dropout, and max-pooling were the most successfully predicted.

\subsection{Attacks Against Distributed Learning}
In the federated learning setting, multiple devices acquire access to the global model that is trained from data that belong to different end users. Furthermore, the parameter server has access to the model updates of each participant either in the form of model parameters or that of loss gradients. In split learning settings, the central server also gains access to the outputs of each participant's intermediate neural network layers. This type of information can be used to mount different types of attacks by actors that are either residing in a central position or even by individual participants. The following subsection presents the types of attacks in distributed settings, as well as their common elements, differences, and assumptions.

\subsubsection{Membership inference attacks}
Nasr et al.~\cite{nasr2018machine} showed that a membership inference attack is more effective than the black-box one, under the assumption that the adversary has some auxiliary knowledge about the training data, i.e., has access to some data from the training dataset, either explicitly or because they are part of a larger set of data that the adversary possesses. The adversary can use the model parameters and the loss gradients as inputs to another model which is trained to distinguish between members and non-members. The white-box attack accuracy with various neural network architectures was up to 75.1\%, however, all target models had a high generalization error.

In the active attack scenario, the attacker, which is also a local participant, alters the gradient updates to perform a gradient ascent instead of descent for the data whose membership is under question. If some other participant uses the data for training, then their local SGD will significantly reduce the gradient of the loss and the change will be reflected in the updated model, allowing the adversary to extract membership information. Attacks from a local active participant reached an attack accuracy of 76.3\% and in general, the active attack accuracy was higher than the passive accuracy in all tested scenarios. However, as the number of participants increases, it has adverse effects on the attack accuracy, which drops significantly after five or more participants. A global active attacker which is in a more favourable position, can isolate the model parameter updates they receive from each participant. Such an active attacker reached an attack accuracy of 92.1\%. 

\subsubsection{Property inference attacks}

Passive property inference requires access to some data that possess the property and some that do not. The attack applies to both federated average and synchronized SGD settings, where each remote participant receives parameter updates from the parameter server after each training round~\cite{melis2019exploiting}. The initial dataset is of the form $\mathcal{D'}=\{(\mathbf{x}, \mathbf{y}, \mathbf{y'})\}$, where $\mathbf{x}$ and $\mathbf{y}$  are the data used for training the distributed model and $\mathbf{y}'$ are the property labels. Every time the local model is updated, the adversary calculates the loss gradients for two batches of data. One batch that has the property in question and one that does not. This allows the construction of a new dataset that consists of gradients and property labels $(\nabla L, \mathbf{y}')$. Once enough labeled data have been gathered, a second model, $f'$, is trained to distinguish between loss gradients of data that have the property versus those that do not. This model is then used to infer whether subsequent model updates were made using data that have the property. The model updates are assumed to be done in batches of data. The attack reaches an attack area under the curve (AUC) score of 98\% and becomes increasingly more successful as the number of epochs increases. Attack accuracy also increases as the fraction of data with the property in question also increases. However, as the number of participants in the distributed model increases, the attack performance decreases significantly.

\subsubsection{Reconstruction attacks}

Some data reconstruction attacks in a federated learning setting use generative models and specifically GANs~\cite{hitaj2017deep, wang2019beyondclass}. When the adversary is one of the participants, they can force the victims to release more information about the class they are interested in reconstructing~\cite{hitaj2017deep}. This attack works as follows: The potential victim has data for a class "A" that the adversary wants to reconstruct. The adversary trains an additional GAN model. After each training round, the adversary uses the target model parameters for the GAN discriminator, whose purpose is to decide whether the input data come from the class "A" or are generated by the generator. The aim of the GAN is to create a generator that is able to generate faithful class "A" samples. In the next training step of the target model, the adversary generates some data using the GAN and labels them as class "B". This forces the target model to learn to discriminate between classes "A" and "B" which in turn improves the GAN training and its ability to generate class "A" representatives.

If the adversary has access to the central parameter server, they have direct access to the model updates of each remote participant. This makes it possible to perform more successful reconstruction attacks~\cite{wang2019beyondclass}. In this case, the GAN discriminator is again using the shared model parameters and learns to distinguish between real and generated data, as well as the identity of the participant. Once the generator is trained, the reconstructed samples are created using an optimization method that minimizes the distance between the real model updates and the updates due to the generated data. Both GAN based methods assume access to some auxiliary data that belong to the victims. However, the former method generates only class representatives.   

In a synchronized SGD setting, an adversary with access to the parameter server has access to the loss gradients of each participant during training. Using the loss gradients is enough to produce a high quality reconstruction of the training data samples, especially when the batch size is small~\cite{zhu2019dlg}. The attack uses a second "dummy" model. Starting with random dummy inputs $x'$ and labels $y'$, the adversary tries to match the dummy model's loss gradients $\nabla_{\theta} \mathcal{J'}$ to the participant's loss gradients $\nabla_{\theta} \mathcal{J}$. This gradient matching is formulated as an optimization task that seeks to find the optimal $x'$ and $y'$ that minimize the gradients' distance:

\begin{equation}
    \label{eq:zhu}
    x^*, y^* = \arg\min_{x',y'} \| \nabla_{\theta} \mathcal{J'}(\mathcal{D'};\theta) - \nabla_{\theta} \mathcal{J}(\mathcal{D};\theta) \|^2
\end{equation}
The minimization problem in Equation~\ref{eq:zhu} is solved using limited memory BFGS (L-BFGS)~\cite{liu1989limited}. 
The size of the training batch is an important factor in the speed of convergence in this attack.


Data reconstruction attacks are also possible during the inference phase in the split learning scenario~\cite{he2019collaborative}. When the local nodes process new data, they perform inference on these initial layers and then send their outputs to the centralized server. In this attack, the adversary is placed in the centralized server and their goal is to try to reconstruct the data used for inference. He et al.~\cite{he2019collaborative} cover a range of scenarios: (i) white-box, where the adversary has access to the initial layers and uses them to reconstruct the images, (ii) black-box where the adversary has no knowledge of the initial layers but can query them and thus recreate the missing layers and (iii) query-free where the adversary cannot query the remote participant and tries to create a substitute model that allows data reconstruction. The latter attack produces the worst results, as expected, since the adversary is the weakest. The split of the layers between the edge device and the centralized server is also affecting the quality of reconstruction. Fewer layers in the edge neural network allow for better reconstruction in the centralized server.

\subsection{Summary of Attacks}
To summarize the attacks proposed against machine learning privacy, Table~\ref{table:attack_summary} presents the 42 papers analyzed in terms of adversarial knowledge, model under attack, attack type, and timing of the attack.

\begin{table*}[t]
\centering
\scriptsize	
\renewcommand{\arraystretch}{1.1}
\caption{Summary of papers on privacy attacks on machine learning systems, including information of their assumptions about adversarial knowledge (black / white-box), the type of model(s) under attack, the attack type, and the timing of the attack (during training or during inference). The transparent circle in the Knowledge column indicates partial white-box attacks.}
\begin{tabular}{@{} l c | c c | c c c c c c c | c c  c c | c c @{} }
\toprule
 Reference & Year & \multicolumn{2}{c}{Knowledge} & \multicolumn{7}{c}{ML Algorithms} & \multicolumn{4}{c}{Attack Type} & \multicolumn{2}{c}{Timing}\\ 
 \midrule
 & & \rotatebox[origin=c]{90}{Black-box} & \rotatebox[origin=c]{90}{White-box} &\rotatebox[origin=c]{90}{Linear regression} &\rotatebox[origin=c]{90}{Logistic regression} &\rotatebox[origin=c]{90}{Decision Trees} & \rotatebox[origin=c]{90}{SVM} & \rotatebox[origin=c]{90}{HMM} & \rotatebox[origin=c]{90}{Neural network} & \rotatebox[origin=c]{90}{GAN / VAE} & \rotatebox[origin=c]{90}{Membership Inference} & \rotatebox[origin=c]{90}{Reconstruction} & \rotatebox[origin=c]{90}{Property Inference} & \rotatebox[origin=c]{90}{Model Extraction} & \rotatebox[origin=c]{90}{Training} & \rotatebox[origin=c]{90}{Inference}  \\
 \midrule
Fredrikson et al.~\cite{fredrikson2014pharma} & 2014  &   & $\bullet$ & $\bullet$ & & & & & & & $\bullet$ & & & & $\bullet$\\  
Fredrikson et al.~\cite{fredrikson2015model} & 2015 & $\bullet$ & $\bullet$  & & & $\bullet$ & & & $\bullet$& & & $\bullet$ & & & & $\bullet$\\
Ateniese et al.~\cite{ateniese2015hacking} & 2015 &  & $\bullet$   & & & & $\bullet$ & $\bullet$& & & & & $\bullet$ & & & $\bullet$ \\
Tramer et al.~\cite{tramer2016stealing}& 2016  & $\bullet$  & $\bullet$  & & $\bullet$ & $\bullet$ & $\bullet$ & &$\bullet$ &  & & & & $\bullet$ & & $\bullet$\\
Wu et al.~\cite{wu2016methodology} & 2016 & $\bullet$ & $\bullet$ &  & $\bullet$ & & & & $\bullet$ & &  & $\bullet$ & & & & $\bullet$\\
Hidano et al.~\cite{hidano2017model} & 2017 & & $\bullet$ & $\bullet$& & & & & & &  & $\bullet$ & & & & $\bullet$\\
Hitaj et al.~\cite{hitaj2017deep}& 2017 &  & $\bullet$  & & & &  & & $\bullet$ & & & $\bullet$ & & & $\bullet$ & \\
Papernot et al.~\cite{papernot2017practical} & 2017 & $\bullet$ & & & & & & &$\bullet$ & & & & & $\bullet$ & & $\bullet$\\
Shokri et al.~\cite{shokri2017membership}  & 2017 & $\bullet$  & & & & & & &$\bullet$ & & $\bullet$ & & &  & & $\bullet$\\
Correia-Silva et al.~\cite{Correia-Silva-IJCNN2018} & 2018 & $\bullet$ & & & & & & &$\bullet$ & & & & & $\bullet$ & & $\bullet$\\
Ganju et al.~\cite{ganju2018property} & 2018 &  & $\bullet$ & & & & & & $\bullet$& & & & $\bullet$ & & & $\bullet$ \\
Oh et al.~\cite{joon2018towards} & 2018 & $\bullet$   & & & & & & & $\bullet$ & & & & & $\bullet$ & & $\bullet$\\
Long et al.~\cite{long2018understanding} & 2018 & $\bullet$  & & & & & & & $\bullet$ & & $\bullet$ & & & & & $\bullet$\\
Rahman et al.~\cite{rahman2018membership} & 2018 &  & $\bullet$ & & & & & & $\bullet$ & & $\bullet$ & & & & & $\bullet$\\
Wang \& Gong~\cite{wang2018stealing} & 2018 &  & $\bullet$   & $\bullet$ & $\bullet$ & & $\bullet$ & & $\bullet$ & & & & &  $\bullet$ & & $\bullet$\\
Yeom et al.~\cite{yeom2018privacy} & 2018 & $\bullet$  & $\circ$  & $\bullet$ & & $\bullet$ & & &$\bullet$ & & & $\bullet$ & &  & & $\bullet$ \\
Carlini et al.~\cite{carlini2019secret} & 2019 & $\bullet$ & & & & & & & $\bullet$ &  &  & $\bullet$ & & & & $\bullet$\\
Hayes et al.~\cite{hayes2019logan} & 2019 & $\bullet$  & $\bullet$  & & & & & & & $\bullet$ & $\bullet$ & & &  & & $\bullet$\\
He et al.~\cite{he2019collaborative} & 2019 & $\bullet$   & $\bullet$  & & & & & & $\bullet$ &  & & $\bullet$ & & & & $\bullet$ \\
Hilprecht et al.~\cite{hilprecht2019monte} & 2019 & $\bullet$  & & & & & & & & $\bullet$ & $\bullet$ & & & & & $\bullet$ \\
Jayaraman \& Evans~\cite{jayaraman2019evaluating} & 2019 &  $\bullet$ & $\bullet$  & & & & & & $\bullet$ & & $\bullet$ & $\bullet$ & & & & $\bullet$\\
Juuti et al.~\cite{juuti2019prada} & 2019 & $\bullet$&   & & & & & & $\bullet$ & &  & & &  $\bullet$&  & $\bullet$\\
Milli et al.~\cite{milli2019model}  & 2019 & $\bullet$& & & & & & & $\bullet$ & & & & & $\bullet$ &  & $\bullet$ \\
Nasr et al.~\cite{nasr2019comprehensive} & 2019 &  & $\bullet$ & & & & & & $\bullet$& & $\bullet$ & & &  & $\bullet$ & \\
Melis et al.~\cite{melis2019exploiting} & 2019 &  & $\bullet$ & & & & & & $\bullet$ & & $\bullet$ & & $\bullet$ & & $\bullet$ &\\
Orekondy et al.~\cite{orekondy2019knockoff} & 2019 & $\bullet$ & & & & & & &$\bullet$ & & & & & $\bullet$ & & $\bullet$\\
Sablayrolles et al.~\cite{sablayrolles2019plmr} & 2019 & & $\circ$ & & & & & & $\bullet$ & & $\bullet$ & & &  & &$\bullet$ \\
Salem et al.~\cite{Salem0HBF019} & 2019 & $\bullet$   & & & & & & & $\bullet$ &  & $\bullet$ & & &  & & $\bullet$\\
Song L. et al.~\cite{song2019privacy} & 2019 & $\bullet$ &  & & & & & & $\bullet$&  & $\bullet$ & & & & & $\bullet$ \\
Truex, et al.~\cite{truex2019demystifying} & 2019 & $\bullet$ & & & $\bullet$ & $\bullet$ & & & $\bullet$ & & $\bullet$ & & & & & $\bullet$\\
 Wang et al.~\cite{wang2019beyondclass} & 2019 & & $\bullet$ & & & & &  & $\bullet$ &  & & $\bullet$ & & & $\bullet$ & \\
Yang et al.~\cite{yang2019neural} & 2019 & $\bullet$ & & & & & & & $\bullet$ &  & & $\bullet$ & & & & $\bullet$ \\
 Zhu et al.~\cite{zhu2019dlg} & 2019 &  & $\bullet$   & & & & & & $\bullet$ &  & & $\bullet$ & & & $\bullet$ & \\
Barbalau et al.~\cite{barbalau2020ripper} & 2020 & $\bullet$ & & & & & & &$\bullet$ & & & & & $\bullet$ & & $\bullet$ \\
 Chandrasekaran et al.~\cite{chandrasekaran2020exploring} & 2020 & $\bullet$ & &  &  &$\bullet$  & $\bullet$ & &$\bullet$ & & & & & $\bullet$ & & $\bullet$\\
Chen et al.~\cite{chen2019gan} & 2020 & $\bullet$ & $\bullet$ & & & & & & & $\bullet$ & $\bullet$ & & & & & $\bullet$ \\
Hishamoto et al.~\cite{hishamoto2020embership} & 2020 & $\bullet$ & & & & & & &$\bullet$ & & $\bullet$ & & & & & $\bullet$\\
Jagielski et al.~\cite{jagielski2020high} & 2020 & $\bullet$ & & & & & & &$\bullet$ & & & & & $\bullet$ & & $\bullet$\\
Krishna et al.~\cite{krishna2020Thieves} & 2020 & $\bullet$ & & & & & & &$\bullet$ & & & & & $\bullet$ & & $\bullet$\\
Pan et al.~\cite{pan2020privacy} & 2020 & & $\bullet$ & & & & & &$\bullet$ & & & $\bullet$ & &  & & $\bullet$\\
Song \& Raghunathan~\cite{song2020infoleak} & 2020 & $\bullet$ & $\bullet$ &  & & & & &$\bullet$ & & $\bullet$ & $\bullet$ & $\bullet$ & & & $\bullet$\\
Zhang et al.~\cite{zhang2020secret} & 2020 & & $\bullet$ & & & & & &$\bullet$ & & & $\bullet$ & &  & & $\bullet$\\
\bottomrule
\end{tabular}

\label{table:attack_summary}
\end{table*}

In terms of model types, 83.3\% of the papers dealt with attacks against neural networks, with decision trees being the second most popular model to attack at 11.9\% (some papers covered attacks against multiple model types). The concept of neural networks groups together both shallow and deep models, as well as multiple architectures, such as convolutional neural networks, recurrent neural networks, while under SVMs we group together both linear and nonlinear versions. 

The most popular attack types are membership inference and reconstruction attacks (35.7\% of the papers, respectively), with model extraction the next most popular (31\%). The majority of the proposed attacks are performed during the inference phase (88\%). Attacks during training are mainly on distributed forms of learning. Black-box and white-box attacks were studied in 66.7\% and 54.8\% of the papers, respectively (some papers covered both settings). In the white-box category, we also include partial white-box attacks.

\begin{figure*}[t]
\centering
\includegraphics[width=1\columnwidth]{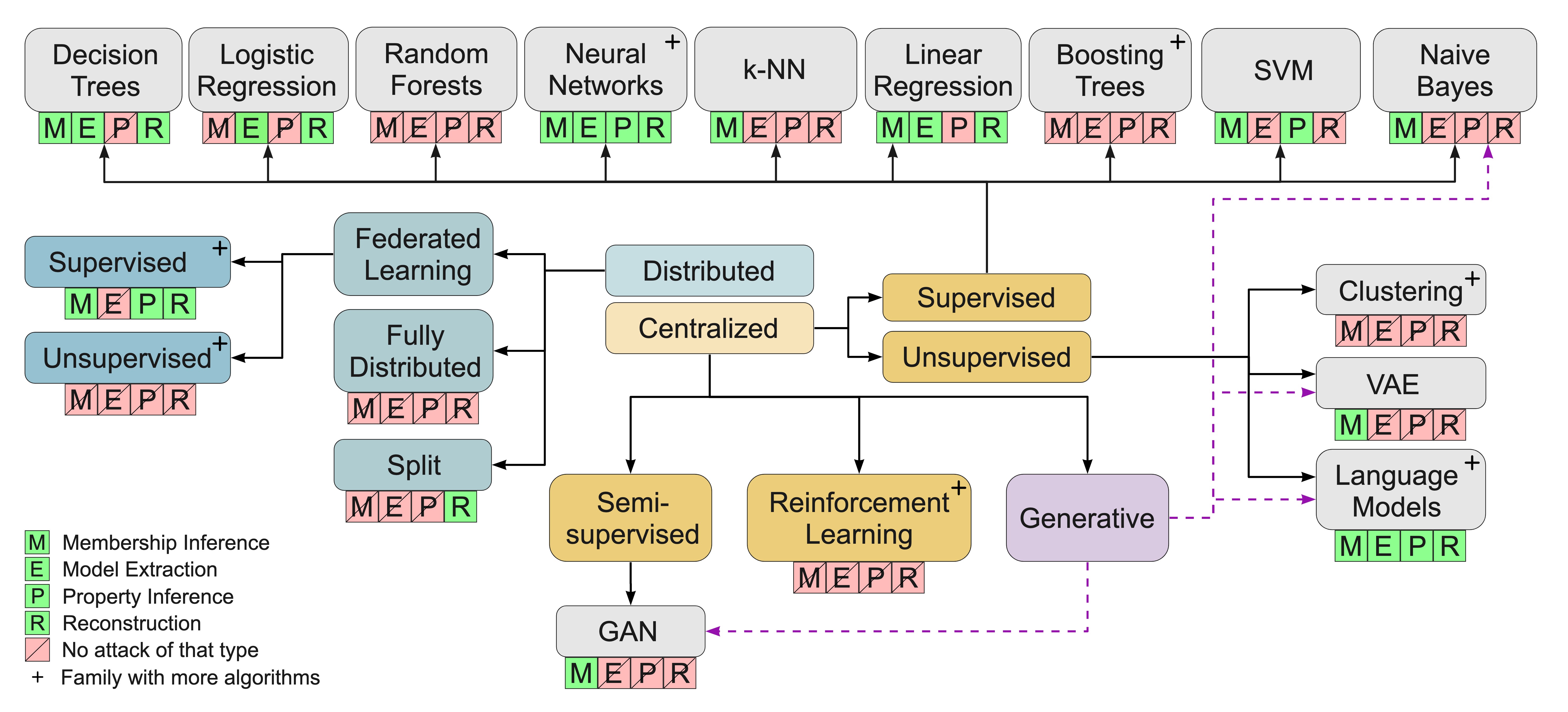}
\caption{Map of attack types per algorithm. The list of algorithm presented is not exhaustive but indicative. Underneath each algorithm or area of machine learning there is an indication of the attacks that have been studied so far. A red box indicates no attack.}
\label{fig:map-attacks}
\end{figure*}

The focus on neural networks in the existing literature as well as the focus on supervised learning is also apparent in Figure~\ref{fig:map-attacks}. The figure depicts types of machine learning algorithms versus the types of attacks that have been studied so far based on the existing literature. The list of algorithms is indicative and not exhaustive, but it contains the most popular ones in terms of research and deployment in real-world systems. Algorithms such as random forests~\cite{breiman2001random} or gradient boosting trees~\cite{chen2016xgboost, ke2017lightgbm} have received little to no focus and the same holds for whole areas of machine learning such as reinforcement learning.  

\begin{figure}[t]
\centering
\includegraphics[width=12cm]{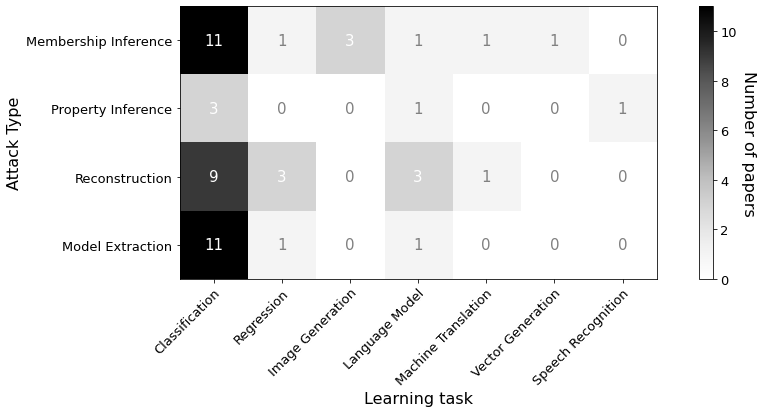}
\caption{Number of papers used against each learning task and attack type. Classification includes both binary and multi-class classification. Darker gray means higher number of papers.}
\label{fig:heatmap}
\end{figure}

Another dimension that is interesting to analyze is the types of learning tasks that have been the target of attacks so far. Figure~\ref{fig:heatmap} presents information about the number of papers in relation to the learning task and the attack type. By learning task, we refer to the task in which the target model is initially trained. As the figure clearly shows, the majority of the attacks are on models that were trained for classification tasks, both binary and multiclass. This is the case across all four attack types. 

\begin{figure}[t]
\centering
\includegraphics[width=12cm]{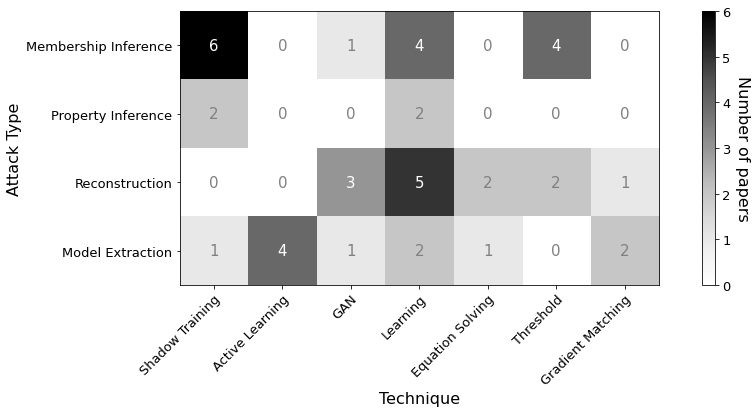}
\caption{Number of papers that used an attacking technique for each attack type. Darker gray means higher number of papers.}
\label{fig:heatmap2}
\end{figure}

While there is a diverse set of reviewed papers, it is possible to discern some high-level patterns in the proposed attacking techniques. Figure~\ref{fig:heatmap2} shows the number of papers in relation to the attacking technique and attack type. Most notably, nine papers used shadow training mainly for membership and property inference attacks. Active learning was quite popular in model extraction attacks and was proposed by four papers. Generative models (mostly GANs) were used in five papers across all attack types and another three papers used gradient matching techniques. It should be noted here that the "Learning" technique includes a number of different approaches, spanning from using model parameters and gradients as inputs to classifiers~\cite{nasr2018machine, melis2019exploiting} to using input-output queries for substitute model creation~\cite{orekondy2019knockoff, Correia-Silva-IJCNN2018, jagielski2020high} and learning classifiers from language models for reconstruction~\cite{pan2020privacy} and property inference~\cite{song2020infoleak}. In "Threshold" based attacks, we categorized the attacks proposed in~\cite{yeom2018privacy} and \cite{sablayrolles2019plmr} and subsequent papers that used them for membership and property inference.

Some attacks may be applicable to multiple learning tasks and datasets, however, this is not the case universally. Dataset size, number of classes, and features might also be factors for the success of certain attacks, especially since most of them are empirical. Table ~\ref{table:datasets_summary1} is a summary of the datasets used in all attack papers along with the data types of their features, the learning task they were used for, and the dataset size. The datasets were used during the training of the target models and in some cases as auxiliary information during the attacks. The table contains 51 unique datasets used across 42 papers, an indication of the variation of different approaches. 

\begin{table*}[!t]
\scriptsize
\renewcommand{\arraystretch}{1.0}
\caption{Summary of datasets used in the papers about privacy attacks on machine learning systems. The size of each dataset is measured by the number of samples unless otherwise indicated. A range in the size column indicates that different papers used different subsets of the dataset.}
\begin{tabular}{p{3cm} p{2cm} p{3.2cm} p{2.5cm} r}
\toprule
\textbf{Name} & \textbf{Data Type} & \textbf{Learning Task} & \textbf{Reference(s)} & \textbf{Size (Samples)}\\
\hline
538 Steak Survey~\cite{538steak} & mixed features & multi-class classification & \cite{fredrikson2015model, tramer2016stealing, chandrasekaran2020exploring, hidano2017model} & 332 \\
AT\&T Faces~\cite{attfaces} & images & multi-class classification & \cite{fredrikson2015model, hitaj2017deep, wang2019beyondclass} & 400 \\
Bank Marketing~\cite{dua2019} & mixed features & multi-class classification & \cite{wang2018stealing} & 45,210 \\
Bitcoin prices & time series & regression & \cite{tramer2016stealing} & 1,076 \\
Book Corpus~\cite{zhu2015aligning} & text & word-level language model & \cite{song2020infoleak} & 14,000 sent. \\
Breast Cancer~\cite{dua2019} & numerical feat. & binary classification & \cite{tramer2016stealing, chandrasekaran2020exploring, long2018understanding} & 699 \\
Caltech 256~\cite{griffin2007caltech} & images & multi-class classification & \cite{orekondy2019knockoff} & 30,607 \\
Caltech birds~\cite{wah2011caltech} & images & multi-class classification & \cite{orekondy2019knockoff} & 6,033 \\
CelebA~\cite{liu2017oblivious} & images & binary classification & \cite{ganju2018property, chen2019gan, yang2019neural, zhang2020secret, barbalau2020ripper} & 20-202,599 \\
CIFAR-10~\cite{krizhevsky2009learning} & images & image generation, multi-class classification & \cite{hayes2019logan, rahman2018membership, Salem0HBF019, shokri2017membership, song2019privacy, yeom2018privacy, sablayrolles2019plmr, he2019collaborative, truex2019demystifying, hilprecht2019monte, milli2019model, jagielski2020high, yang2019neural, barbalau2020ripper} & 60,000 \\
CIFAR-100~\cite{krizhevsky2009learning} & images & multi-class classification & \cite{nasr2019comprehensive, Salem0HBF019, shokri2017membership, yeom2018privacy, zhu2019dlg, jayaraman2019evaluating, barbalau2020ripper} & 60,000 \\
CLiPS stylometry~\cite{verhoeven2014clips} & text & binary classification & \cite{melis2019exploiting} & 1,412 reviews \\
Chest X-ray~\cite{wang2017chest} &images & multi-class classification & \cite{zhang2020secret} & 10,000\\
Diabetes~\cite{dua2019} & time series & binary class., regression & \cite{tramer2016stealing, wang2018stealing, chandrasekaran2020exploring} & 768 \\
Diabetic ret.~\cite{diabeticretinopathydata} & images & image generation & \cite{hayes2019logan, orekondy2019knockoff} & 88,702 \\
Enron emails & text & char-level language model & \cite{carlini2019secret} & - \\  
Eyedata~\cite{scheetz2006regulation} & numerical feat.& regression & \cite{yeom2018privacy} & 120 \\
FaceScrub~\cite{ng2014data} & images & binary classification & \cite{melis2019exploiting, yang2019neural} & 18,809-48,579 \\ 
Fashion-MNIST~\cite{xiao2017fashion} & images & multi-class classification & \cite{song2019privacy, hilprecht2019monte, jagielski2020high, barbalau2020ripper} & 60,000 \\
Foursquare~\cite{yang2015nationtelescope} & mixed features & binary classification & \cite{melis2019exploiting, shokri2017membership, Salem0HBF019} & 528,878 \\
Geog. Orig. Music~\cite{dua2019} & numerical feat. & regression & \cite{wang2018stealing} & 1,059 \\
German Credit~\cite{dua2019} & mixed features & binary classification & \cite{tramer2016stealing} & 1,000 \\
GSS marital survey~\cite{gssmarital} & mixed features & multi-class classification & \cite{fredrikson2015model, tramer2016stealing, chandrasekaran2020exploring} & 16127 \\
GTSRB~\cite{stallkamp2011german} & images & multi-class classification & \cite{juuti2019prada, papernot2017practical} & 51839\\
HW Perf. Counters (private) & numerical feat. & binary classification & \cite{ganju2018property} & 36,000 \\
Imagenet~\cite{imagenet_cvpr09} & images & multi-class classification & \cite{joon2018towards, sablayrolles2019plmr, jagielski2020high, barbalau2020ripper} & 14,000,000 \\
Instagram~\cite{backes2017walk2friends} & location data & vector generation & \cite{chen2019gan} & - \\
Iris~\cite{fisher1936use} & numerical feat. & multi-class classification & \cite{tramer2016stealing, chandrasekaran2020exploring} & 150 \\
IWPC~\cite{international2009estimation} & mixed features & regression & \cite{fredrikson2014pharma, yeom2018privacy} & 3497 \\
IWSLT Eng-Vietnamese~\cite{luong2015iwslt15} & text & neural machine translation & \cite{carlini2019secret} & - \\
LFW~\cite{huang2008labeled}& images & image generation & \cite{hayes2019logan, melis2019exploiting, zhu2019dlg} & 13233 \\ 
Madelon~\cite{dua2019} & mixed features & multi-class classification & \cite{wang2018stealing} & 4,400 \\
MIMIC-III~\cite{johnson2016mimic} & binary features & record generation & \cite{chen2019gan} & 41,307 \\
Movielens 1M~\cite{movielens} & numerical feat. & regression & \cite{hidano2017model} & 1,000,000 \\
MNIST~\cite{yann1998mnist} & images & multi-class classification & \cite{ganju2018property, hitaj2017deep, joon2018towards, rahman2018membership, Salem0HBF019, shokri2017membership, tramer2016stealing, yeom2018privacy, zhu2019dlg, he2019collaborative, truex2019demystifying, hilprecht2019monte, wang2019beyondclass, juuti2019prada, papernot2017practical, milli2019model, chandrasekaran2020exploring, jagielski2020high, yang2019neural, zhang2020secret, long2018understanding} & 70,000 \\
Mushrooms~\cite{dua2019} & categorical feat. & binary classification & \cite{tramer2016stealing, chandrasekaran2020exploring} & 8,124 \\
Netflix~\cite{netflixdataset} & binary features & binary classification & \cite{yeom2018privacy} & 2,416 \\
Netflows (private) & network data & binary classification & \cite{ateniese2015hacking} & - \\
PTB~\cite{marcus1993building} & text & char-level language model & \cite{carlini2019secret} & 5 MB \\
PiPA~\cite{zhang2015beyond} & images & binary classification & \cite{melis2019exploiting} & 18,000 \\
Purchase-100~\cite{purchase100} & binary features & multi-class classification & \cite{nasr2019comprehensive, shokri2017membership, truex2019demystifying, jayaraman2019evaluating} & 197,324 \\
SVHN~\cite{netzer2011reading} & images & multi-class classification & \cite{zhu2019dlg, jagielski2020high} & 60,000 \\
TED talks~\cite{tedtalksiwslt} & text & machine translation & \cite{carlini2019secret} & 100,000 pairs \\

Texas-100~\cite{texashealthdata} & mixed features & multi-class classification & \cite{nasr2019comprehensive, shokri2017membership} & 67,330 \\ 
UJIndoor~\cite{dua2019} & mixed features & regression & \cite{wang2018stealing} & 19,937 \\
UCI / Adult~\cite{dua2019} & various & binary classification & \cite{ganju2018property, Salem0HBF019, shokri2017membership, tramer2016stealing, truex2019demystifying, chandrasekaran2020exploring, long2018understanding} & 48,842 \\
Voxforge~\cite{voxforgedata} & audio & speech recognition & \cite{ateniese2015hacking} & 11,137 rec. \\
Wikipedia~\cite{mahoney2009wiki} & text & language model & \cite{song2020infoleak} & 150,000 articles \\
Wikitext-103~\cite{merity2016pointer} & text & word-level language model & \cite{carlini2019secret, krishna2020Thieves} & 500 MB \\ 
Yale-Face~\cite{georghiades2001few} & images & multi-class classification & \cite{song2019privacy} & 2,414 \\
Yelp reviews~\cite{yelpopendataset} & text & binary classification & \cite{melis2019exploiting} & 16-40,000 \\
\bottomrule
\end{tabular}

\label{table:datasets_summary1}
\end{table*}

This high variation is both a blessing and a curse. On the one hand, it is highly desirable to use multiple types of datasets to test different hypotheses and the majority of the reviewed research follows that approach. On the other hand, these many options make it harder to compare methods. As it is evident from Table~\ref{table:datasets_summary1}, some of the datasets are quite popular. MNIST, CIFAR-10, CIFAR-100, and UCI Adult have been used by more than six papers, while 26 datasets have been used by only one paper.

The number of model parameters varies based on the model, task and datasets used in the experiments. As it can be seen in Table~\ref{table:datasets_summary1}, most datasets are not extremely large, hence the models under attack are not extremely large. Given that most papers deal with neural networks, this might indicate that most attacks focused on smaller datasets and models which might not be representative of realistic scenarios. However, privacy attacks do not necessarily have to target large models with extreme amounts of data; and neural networks, however popular, are not necessarily the most used models in the "real world".

\section{Defending Machine Learning Privacy}
\label{sec:defenses}
Leaking personal information such as medical records or credit card numbers is usually an undesirable situation. The purpose of studying attacks against machine learning models is to be able to explore the limitations and assumptions of machine learning and to anticipate the adversaries' actions. Most of the analyzed papers propose and test mitigations to counter their attacks. In the next subsections, we present the various defences proposed in several papers organized by the type of attack they attempt to defend against.


\subsection{Defenses Against Membership Inference Attacks}
The most prominent defense against membership inference attacks is Differential Privacy (DP), which provides a guarantee on the impact that single data records have on the output of an algorithm or a model. However, other defenses have been tested empirically and are also presented in the following subsections.

\subsubsection{Differential Privacy}
Differential privacy started as a privacy definition for data analysis and it is based on the idea of "learning nothing about an individual while learning useful information about a population"~\cite{Dwork2013}. Its definition is based on the notion that if two databases differ only by one record and are used by the same algorithm (or mechanism), the output of that algorithm should be similar. More formally,

\begin{definition}[($\epsilon,\delta$)-Differential Privacy]
A randomized mechanism $\mathcal{M}$ with domain $\mathcal{R}$ and output $\mathcal{S}$ is ($\epsilon,\delta$)-differentially private if for any adjacent inputs $D, D' \in \mathcal{R}$ and for any subsets of outputs $\mathcal{S}$ it holds that:

\begin{equation}
  Pr[\mathcal{M}(D) \in \mathcal{S} ] \leq e^{\epsilon} Pr[\mathcal{M}(D') \in \mathcal{S} ] +\delta  
\end{equation}
where $\epsilon$ is the privacy budget and $\delta$ is the failure probability. 
\end{definition}
The original definition of DP did not include $\delta$ which was introduced as a relaxation that allows some outputs not to be bounded by $e^{\epsilon}$. 

The usual application of DP is to add Laplacian or Gaussian noise to the output of a query or function over the database. The amount of noise is relevant to the \textit{sensitivity} which gives an upper bound on how much we must perturb the output of the mechanism to preserve privacy~\cite{Dwork2013}:

\begin{definition}
$l_{1}$ (or $l_{2}$)-Sensitivity of a function $f$ is defined as  
\begin{equation}
  \Delta f = \underset{D, D', \|D-D'\|=1}{\max} \|f(D) - f(D')\|  
\end{equation}
\end{definition}

where $\|.\|$ is the $l_1$ or the $l_2$-norm and the max is calculated over all possible inputs $D, D'$. 

From a machine learning perspective, $D$ and $D'$ are two datasets that differ by one training sample and the randomized mechanism $\mathcal{M}$ is the machine learning training algorithm. In deep learning, the noise is added at the gradient calculation step. Because it is necessary to bound the gradient norm, gradient clipping is also applied~\cite{Abadi2016}.

Differential privacy offers a trade-off between privacy protection and utility or model accuracy. Evaluation of differentially private machine learning models against membership inference attacks concluded that the models could offer privacy protection only when they considerably sacrifice their utility~\cite{rahman2018membership, jayaraman2019evaluating}. Jayaraman et al.~\cite{jayaraman2019evaluating} evaluated several relaxations of DP in both logistic regression and neural network models against membership inference attacks. They showed that these relaxations have an impact on the utility-privacy trade-off. While they reduce the required added noise, they also increase the privacy leakage.

Distributed learning scenarios require additional considerations regarding differential privacy. In a centralized model, the focus is on sample level DP, i.e., on protecting privacy at the individual data point level. In a federated learning setting where there are multiple participants, we not only care about the individual training data points they use, but also about ensuring privacy at the participant level. A proposal which applies DP at the participant level was introduced by McMahan et al.~\cite{brendan2018learning} however, it requires a large number of participants. When it was tested with a number as low as 30, the method was deemed unsuccessful~\cite{melis2019exploiting}.

\subsubsection{Regularization}

Regularization techniques in machine learning aim to reduce overfitting and increase model generalization performance. Dropout~\cite{srivastava2014dropout} is a form of regularization that randomly drops a predefined percentage of neural network units during training. Given that black-box membership inference attacks are connected to overfitting, it is a sensible approach to this type of attack and multiple papers have proposed it as a defense with varying levels of success~\cite{hayes2019logan, melis2019exploiting, Salem0HBF019, shokri2017membership, song2019privacy}. Another form of regularization uses techniques that combine multiple models that are trained separately. One of those methods, model stacking, was tested in~\cite{Salem0HBF019} and produced positive results against membership inference. An advantage of model stacking or similar techniques is that they are model agnostic and do not require that the target model is a neural network.

\subsubsection{Prediction vector tampering} As many models assume access to the prediction vector during inference, one of the countermeasures proposed was the restriction of the output to the top k classes or predictions of a model~\cite{shokri2017membership}. However, this restriction, even in the strictest form (outputting only the class label) did not seem to fully mitigate membership inference attacks, since information leaks can still happen due to model misclassifications. Another option is to lower the precision of the prediction vector, which leads to less information leakage~\cite{shokri2017membership}. Adding noise to the output vector also affected membership inference attacks~\cite{jia2019memguard}. 

\subsection{Defenses Against Reconstruction Attacks}
Reconstruction attacks often require access to the loss gradients during training. Most of the defences against reconstruction attacks propose techniques that affect the information retrieved from these gradients. Setting all loss gradients which are below a certain threshold to zero, was proposed as a defence against reconstruction attacks in deep learning. This technique proved quite effective with as little as 20\% of the gradients set to zero and with negligible effects on model performance~\cite{zhu2019dlg}. On the other hand, performing quantization or using half-precision floating points for neural network weights did not seem to deter the attacks in~\cite{carlini2019secret} and ~\cite{zhu2019dlg}, respectively.



\subsection{Defenses Against Property Inference Attacks}

Differential privacy is designed to provide privacy guarantees in membership inference attack scenarios and it does not seem to offer protection against property inference attacks~\cite{ateniese2015hacking}. In addition to DP, Melis et al.~\cite{melis2019exploiting} explored other defenses against property inference attacks. Regularization (dropout) had an adverse effect and actually made the attacks stronger. Since the attacks in~\cite{melis2019exploiting} were performed in a collaborative setting, the authors tested the proposal in~\cite{shokri2015privacy}, which is to share fewer gradients between training participants. Although sharing less information made the attacks less effective, it did not alleviate them completely.

\subsection{Defenses Against Model Extraction Attacks}
Model extraction attacks usually require that the attacker performs a number of queries on the target model. The goal of the proposed defenses so far has been the detection of these queries. This contrasts with the previously presented defences that mainly try to prevent attacks.

\subsubsection{Protecting against DNN Model Stealing Attacks (PRADA)} Detecting model stealing attacks based on the model queries that are used by the adversary was proposed by Juuti et al.~\cite{juuti2019prada}. The detection is based on the assumption that model queries that try to explore decision boundaries will have a different distribution than the normal ones. While the detection was successful, the authors noted that it is possible to be evaded if the adversary adapts their strategy.
\subsubsection{Membership inference} The idea of using membership inference to defend against model extraction was studied by Krishna et al.~\cite{krishna2020Thieves}. It is based on the premise that using membership inference, the model owner can distinguish between legitimate user queries and nonsensical ones whose only purpose is to extract the model. The authors note that this type of defence has limitations such as potentially flagging legitimate but out-of-distribution queries made by legitimate users, but more importantly that they can be evaded by adversaries that make adaptive queries.

\section{Discussion}
\label{sec:discussion}
Attacks on machine learning privacy have been increasingly brought to light. However, we are still at an exploratory stage. Many of the attacks are applicable only under specific sets of assumptions or do not scale to larger training data sets, number of classes,  number of participants, etc. The attacks will keep improving and to successfully defend against them, the community needs to answer fundamental questions about why they are possible in the first place. While progress has been made in the theoretical aspects of some of the attacks, there is still a long way to go to achieve a better theoretical understanding of privacy leaks in machine learning.

As much as we need answers about why leaks happen at a theoretical level, we also need to know how well privacy attacks work on real deployed systems. Adversarial attacks on realistic systems bring to light the issue of additional constraints that need to be in place for the attacks to work. When creating glasses that can fool a face recognition system, Sharif et al.~\cite{sharif2016accessorize}, they had to pose constraints that had to do with physical realizations, e.g., that the color of the glasses should be printable. In privacy-related attacks, the most realistic cases come from the model extraction area, where attacks against MLaaS systems have been demonstrated in multiple papers. For the majority of other attacks, it is certainly an open question of how well they would perform on deployed models and what kind of additional requirements need to be in place for them to succeed.

At the same time, the main research focus up to now has been supervised learning. Even within supervised learning, there are areas and learning tasks that have been largely unexplored, and there are few attacks reported on popular algorithms such as random forests or gradient boosting trees despite their wide application. In unsupervised and semi-supervised learning, the focus is mainly on generative models and only just recently, papers started exploring areas such as representation learning and language models. Some attacks on image classifiers do not transfer that well to natural language processing tasks~\cite{hishamoto2020embership} while others do, but may require different sets of assumptions and design considerations~\cite{pan2020privacy}.     

Beyond expanding the focus on different learning tasks, there is the question of datasets. The impact of datasets on the attack success has been demonstrated by several papers. Yet, currently, we lack a common approach as to which datasets are best suited to evaluate privacy attacks, or constitute the minimum requirement for a successful attack. Several questions are worth considering: do we need standardized datasets and if yes, how do we go about and create them? Are all data worth protecting and if some are more interesting than others, should we not be testing attacks beyond popular image datasets?

Finally, as we strive to understand the privacy implications of machine learning, we also realize that several research areas are connected and affect each other. We know, for instance, that adversarial training affects membership inference~\cite{shokri2019privacy} and that model censoring can still leak private attributes~\cite{Song2020Overlearning}. Property inference attacks can deduce properties of the training dataset that were not specifically encoded or were not necessarily correlated to the learning task. This can be understood as a form of bias detection, which means that relevant literature in the area of model fairness should be reviewed as potentially complementary. Furthermore, while deep learning models are considered black-boxes in terms of explainability, work that sheds light on what kind of data make neurons activate~\cite{yosinski2015understanding, nguyen2016synth} can be relevant to discovering information about the training dataset and can therefore lead to privacy leaks. All these are examples of potential inter-dependencies between different areas of machine learning research, therefore, a better understanding of privacy attacks calls for an interdisciplinary approach. 

\section{Conclusion}
\label{sec:conclusion}
As machine learning becomes ubiquitous, the scientific community becomes increasingly interested in its impact and side-effects in terms of security, privacy, fairness, and explainability. This survey conducted a comprehensive study of the state-of-the-art privacy-related attacks and proposed a threat model and a unifying taxonomy of the different types of attacks based on their characteristics. An in-depth examination of the current state of the art research allowed us to perform a detailed analysis which revealed common design patterns and differences between them. 

Several open problems that merit further research were identified. First, our analysis revealed a somewhat narrow focus of the research conducted so far, which is dominated by attacks on deep learning models. We believe that there are several popular algorithms and models in terms of real-world deployment and applicability that merit a closer examination. Second, a thorough theoretical understanding of the reasons behind privacy leaks is still underdeveloped and this affects both the proposed defensive measures and our understanding of the limitations of privacy attacks. Experimental studies on factors that affect privacy leaks have provided useful insights so far. However, in total, there are very few works that test attacks in realistic conditions in terms of dataset size and deployment. Finally, examining the impact of other adjacent study areas such as security, explainability, and fairness is also a topic that calls for further exploration. Even though it may not be possible to construct and deploy models that are fully private against all types of adversaries, understanding the inter-dependencies that affect privacy will help make more informed decisions. 

While the community is still in an exploratory mode regarding privacy leaks of machine learning systems, we hope that this survey will provide the necessary background to both the interested readers as well as the researchers that wish to work on this topic.

\begin{acks}
This work was partially supported by Avast Software and the OP RDE funded project Research Center for Informatics No.: CZ.02.1.01/0.0./0.0./16\_019/0000765. 
\end{acks}

\bibliographystyle{ACM-Reference-Format}
\bibliography{references}










\end{document}